\documentclass[aps,reprint,nofootinbib,nobibnotes,notitlepage,superscriptaddress,onecolumn,prd,
 amsmath,amssymb
]{revtex4-2}

\usepackage[caption=false]{subfig}
\usepackage{braket}
\usepackage{float}
\usepackage{lipsum}
\usepackage{booktabs}
\usepackage{graphicx}
\usepackage{dcolumn}
\usepackage{bm}
\usepackage{natbib}
\usepackage{hyperref}
\hypersetup{
	colorlinks = true,
    linkcolor = Red,
    urlcolor  = Red,
    citecolor = Red
}
\usepackage{subfig}

\usepackage{microtype}

\usepackage{verbatim}
\usepackage[amssymb]{SIunits}
\usepackage{tabularx}
\usepackage{longtable}
\usepackage[dvipsnames]{xcolor}
\usepackage{wasysym}
\usepackage[export]{adjustbox}
\usepackage{multirow}
\usepackage{tikz}

\def\be{\begin{equation}}
\def\ee{\end{equation}}
\usepackage{enumitem}

\usepackage{color}
\definecolor{darkgreen}{RGB}{0,120,0}

\definecolor{darkgreen}{RGB}{0,120,0}

\newcommand{\av}[1]{\left\langle{#1}\right\rangle}

\newcommand{\F}{\mathcal{F}}
\newcommand{\G}{\mathcal{G}}

\def\beq{\begin{eqnarray}}
\def\eeq{\end{eqnarray}}
\let\vec\mathbf

\usepackage{empheq}

\usepackage{booktabs}

\begin{document}

\title{Searching for Unparticles with the Cosmic Microwave Background}

\author{Oliver~H.\,E.~Philcox}
\email{ohep2@cantab.ac.uk}
\affiliation{Leinweber Institute for Theoretical Physics at Stanford, 382 Via Pueblo, Stanford, CA 94305, USA}
\affiliation{Kavli Institute for Particle Astrophysics and Cosmology, 382 Via Pueblo, Stanford, CA 94305, USA}
\author{Guilherme~L.~Pimentel}
\email{guilherme.leitepimentel@sns.it}
\affiliation{Scuola Normale Superiore and INFN, Piazza dei Cavalieri 7, 56126, Pisa, Italy}
\author{Chen Yang}
\email{chen.yang@sns.it}
\affiliation{Scuola Normale Superiore and INFN, Piazza dei Cavalieri 7, 56126, Pisa, Italy}

\begin{abstract} 
\noindent
Multi-field models of inflation typically assume that interactions between particles are weak and can be treated perturbatively. Strongly-coupled models provide an intriguing alternative and may offer novel inflationary phenomenology. In this work, we study the ``unparticle'' scenario, where the inflaton is weakly mixed with a strongly-coupled sector, specified by a (gapless) conformal field theory. For certain choices of the conformal scaling dimension, $\Delta$, the exchange of unparticles leads to distinctive non-Gaussian features in the primordial curvature distribution, including bispectra with enhanced squeezed limits and oscillations close to the equilateral regime. Efficiently analyzing these models using Cosmic Microwave Background (CMB) data is a challenge since (a) they are defined over a wide range of $\Delta$, (b) the novel phenomenology is not restricted to squeezed limits, (c) the bispectra are non-factorizable in momenta, (d) the signatures are often highly degenerate with single-field self-interactions. Here, we overcome these limitations using a library of tools, including recently developed neural-network factorization schemes, principal component analyses (with carefully-chosen bases), and optimal CMB estimators. Our combined pipeline condenses 161 non-separable unparticle bispectra into just 7 factorizable forms, with negligible loss of signal-to-noise. We apply the model to the latest temperature and polarization data from \textit{Planck}, asking two key questions: (1) can we detect unparticles? (2) can we distinguish between unparticle bispectra and single-field self-interactions? Across $1\leq \Delta\leq 9$, we find a maximal signal-to-noise of $1.2\sigma$ (or $1.7\sigma$ in an analysis marginalized over self-interactions), implying no evidence for new physics. We also place the first CMB constraints on the modified consistency-condition-satisfying orthogonal bispectrum typically used in galaxy survey analyses with $f^{\rm orth^*}_{\rm NL} = -12\pm12$. While many unparticle models are very degenerate with the single-field shapes (and highly correlated amongst themselves), certain values of $\Delta$ (close to half-integers) have very different shapes, offering an intriguing discovery channel for future experiments. The methods developed herein can be directly applied to other classes of strongly-coupled sectors (and beyond), motivating the exploration of models beyond the standard weakly-coupled paradigm.
\end{abstract}

\maketitle

\section{Introduction}\label{sec: intro}
\noindent The primordial bispectrum is an important probe of the underlying dynamics of inflation. It is instructive to consider the parallels between the bispectrum in cosmology and scattering events in terrestrial collider physics: both probe the microscopic laws responsible for the low-energy interactions of light external particles---curvature perturbations in inflation or partons in the LHC--with high-energy sectors that cannot be observed directly. In both scenarios, our inference of fundamental physics is limited by our imaginations: restricting our attention to simple bispectrum templates, such as the local, orthogonal and equilateral shapes, can potentially lead to a loss of novel microscopic information about inflation. In this paper, we explore a scenario of an ``unparticle" mediating an interaction among the curvature fluctuations during inflation. This is a very simple model of an underlying \textit{strongly coupled sector} in inflation, which contrasts with the conventional weakly coupled ``quasi-single field" templates that have been analyzed in detail in recent years; e.g.\,\cite{Maldacena:2002vr,Chen:2009we,Chen:2009zp,Chen:2010xka,Baumann:2011nk,Assassi:2012zq,Noumi:2012vr,Arkani-Hamed:2015bza,Arkani-Hamed:2018kmz,Cui:2021iie,Reece:2022soh,Qin:2022lva,Pimentel:2022fsc,Wang:2022eop,Wang:2025qww,deRham:2025mjh,Cai:2025kcd,Wang:2025qfh}.  

\begin{figure}[!t]
    \centering
    \includegraphics[width=0.9\linewidth]{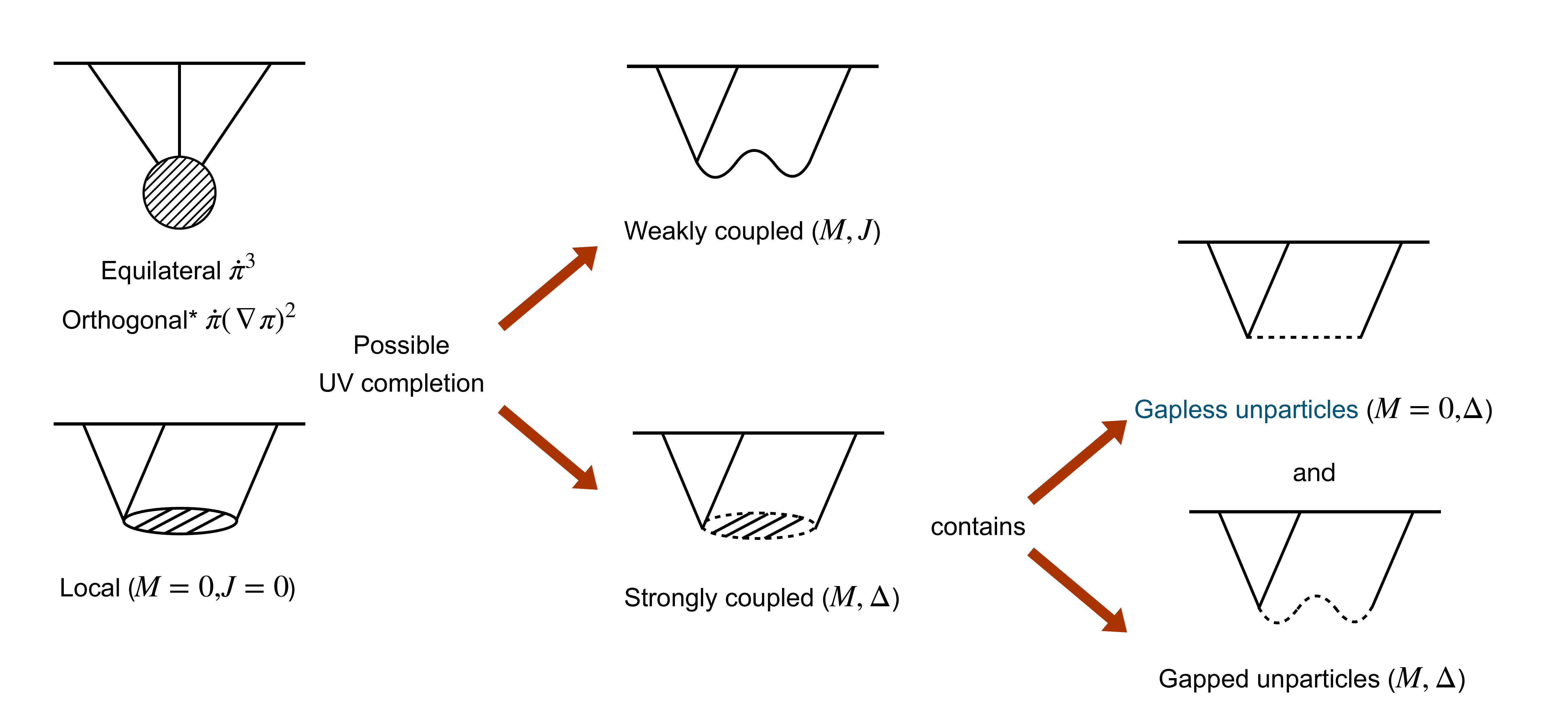}
    \caption{\textbf{Types of inflationary bispectra}. The EFT of inflation provides a basis for describing self-interactions of curvature perturbations. Often, one assumes that the microscopic physics underlying this self-interaction is weakly coupled, sourced by a mediator particle of generic mass and spin. In the massless limit, where the mediator cannot be integrated out, this results in local non-Gaussianity, whilst the heavy-mass case leads to equilateral signatures. Alternatively, the microscopic dynamics can also arise from the exchange of an operator in a strongly coupled sector. In this case (which has rarely been considered in the literature) we can parameterize the correlation function of the strongly coupled theory by its mass gap and the scaling dimension of the operator. In this work, we focus on a gapless model known as the unparticle scenario, which is determined only by its scaling dimension, $\Delta$.}
    \label{fig: templates}
\end{figure}

To begin, let us quickly review the theoretical framework from which we build our templates \cite{Cheung:2007st,Baumann:2009ds,Senatore:2010wk,Green:2013rd,Lee:2016vti,Pimentel:2025rds,Yang:2025apy,Jiang:2025mlm}. If primordial fluctuations really came from inflation, we can organize the dynamics of the fluctuations using an Effective Field Theory---the EFT of inflation. In this case, self-interactions of the curvature perturbation are encoded in a handful of local operators, and can be searched for using the well-known equilateral and orthogonal shapes of the bispectrum.\footnote{Formally, we require the modified orthogonal template introduced in \cite{Senatore:2009gt} to fully characterize self-interactions, since the conventional form has an unphysical squeezed limit. In practice, the standard orthogonal template is often sufficient however.} The EFT of inflation is similar to the non-linear sigma model in particle physics. In particular, being an EFT, it requires new physics at high energies to explain its inherently non-linear dynamics.\footnote{In the standard slow-roll scenario, a single scalar field is responsible for the background perturbations, and its mixing with gravity produces the non-linear structure of the EFT. Going beyond this minimal scenario requires new physics in the form of additional fields.}
This leads us to consider candidate microscopic/ultraviolet (UV) completions of the EFT. The simplest possibility---where most models of inflation fall within---is of a UV completion with a linear sigma model (LSM), such as slow-roll inflation. In this case, the inflaton has a classical trajectory that determines the background geometry, and its small fluctuations source curvature perturbations. 
Another natural, but much less explored, UV completion is to regard the spectator fields as a composite particle in a strongly coupled theory.\footnote{Such a scenario would be suggestive of the inflaton itself also being a composite of more fundamental fields. It'd be interesting to investigate the phenomenological consequences of such a possibility.} In particle physics, examples of this form include Quantum Chromodynamics (QCD) and Composite scenarios for the Higgs. We illustrate these different UV completion scenarios in Fig.~\ref{fig: templates}.

Amongst the strongly coupled models, the most peculiar signal comes from those the large anomalous dimensions of the composite fields. Anomalous dimensions are ubiquitous in quantum mechanics---they are a modification of the naive scaling behavior of correlations among particles, due to interactions either among themselves or with their environment. In practice, the usual Newtonian/Coulomb potential among particles can be modified with a different fall-off rate. The modified power-law decay is the anomalous dimension. If interactions are small, anomalous dimensions will be very small, so large anomalous dimensions, like the ones we study here, are tied to strongly coupled dynamics. To obtain a significant anomalous dimension in the composites, from the classification of infrared (IR) behaviors of quantum field theories (QFTs) \cite{Rychkov:2016iqz}, the IR theory has to be scale-invariant.\footnote{This scale-invariance is of the {\it four-dimensional} theory coupled to inflation. It is a larger symmetry than the one we usually require for inflation. The scale-invariance of primordial perturbations is a consequence of de Sitter symmetry, and tied to the lightness of the inflaton in Hubble units.} 
This scale-invariant field is often referred to as an unparticle \cite{Georgi:2007ek,Grinstein:2008qk,Strassler:2008bv}, and can be described as operators of a conformal field theory (CFT). 
The behaviors of such models in the squeezed and collapsed configurations were first discussed in \cite{Green:2013rd}, while the bispectrum and trispectrum for generic kinematics were studied in \cite{Pimentel:2025rds}. For a model where strong coupling dynamics arises during inflation, see e.g. the ``Dark Walker" from the conformal window of QCD in \cite{Yang:2025apy}. There, some operators act as unparticles during inflation, while the theory opens up gaps and has dark matter candidates after inflation. 
These models exhibit a distinctive inflationary phenomenology. In particular, the interactions generate coherent correlations between modes, leading to oscillatory features in the bispectrum that extend beyond the squeezed limit.

If such sectors are present during inflation, how can we detect them? In general, searching for novel inflationary signatures is a difficult task, given the coupled complexities of predicting the correlators in inflationary space, evolving them to late-times (and potentially non-linear scales), and efficiently probing them using data from the Cosmic Microwave Background (CMB) or Large Scale Structure (LSS). Here, we will focus on the first observable, which currently provides the strongest constraints on most primordial signatures (except for some oscillatory features) and is simpler to model, due to the small amplitude of fluctuations at the redshifts corresponding to the primary CMB signal. 

\begin{figure}[!t]
    \centering
    \includegraphics[width=0.9\linewidth]{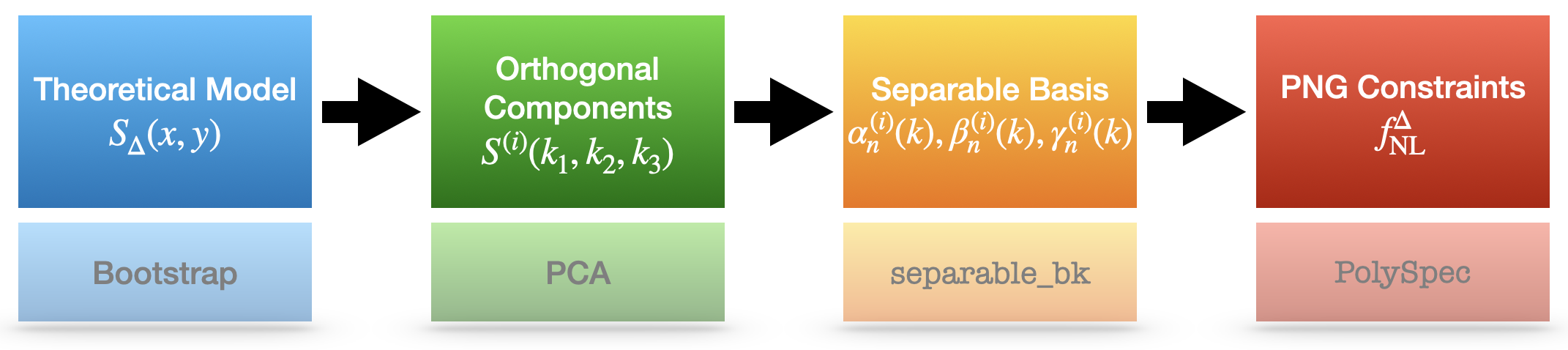}
    \caption{\textbf{Schematic outline of this work}. Given a set of primordial bispectrum templates, computed analytically using bootstrap techniques, we use linear optimization (a principal component analysis; PCA) to obtain a small set of basis templates that describe a wide span of models. Using the techniques of \citep{Philcox:2025bbo}, these are expanded in factorizable form using machine learning techniques, allowing them to be integrated in KSW-type CMB estimators. Finally, we use the \textsc{PolySpec} package to place constraints on the underlying template amplitudes, $f_{\rm NL}^{\Delta}$. Though we focus on unparticle templates in this work, this outline could be applied to any suite of models.}
    \label{fig: outline}
\end{figure}

Due to the high-dimensionality of contemporary CMB data (with $N_{\rm pix} \mathcal{O}(10^{7})$ pixels in \textit{Planck}), we require sophisticated techniques to robustly extract the small inflationary bispectrum signatures. The standard work-horse is the Komatsu-Spergel-Wandelt (KSW) approach, which invokes factorizability of the underlying inflationary bispectrum to reduce the computation time of a bispectrum estimator from $\mathcal{O}(N_{\rm pix}^3)$ to $\mathcal{O}(N_{\rm pix}\log N_{\rm pix})$. Many models, such as the unparticle model introduced above, are not inherently factorizable, thus we must proceed by an alternate (and usually approximate) path, such as measuring the bispectrum in a set of Legendre modes, Fourier-modes, or bins \citep[e.g.,][]{2009PhRvD..80d3510F,2011arXiv1105.2791F,Bucher:2015ura,Munchmeyer:2014nqa,2011MNRAS.417....2S,Bucher:2009nm,Sohn:2023fte}. Here, we adopt a different approach, which is sketched in Fig.\,\ref{fig: outline}. utilizing the machine learning framework introduced in \citep{Philcox:2025bbo} to obtain a highly accurate approximation to the target bispectrum of interest using a set of neural network basis functions. Unlike the conventional equilateral and orthogonal templates, our models depend on an \textit{a priori} unknown parameter, the conformal scaling dimension $\Delta$. To efficiently characterize the broad parameter space, we supplement the factorization approach with a Principal Component Analysis (PCA) compression (paying close attention to possible degeneracies with single-field bispectra), and implement the result in the CMB bispectrum code \textsc{PolySpec}. Finally, we apply our results to the latest \textit{Planck} temperature and polarization dataset, which (a) places strong constraints on the unparticle scenario (optionally marginalizing over self-interactions), and (b) serves as a proof-of-concept for the analysis of general strongly-coupled (and non-factorizable) inflationary models.

The remainder of this paper is as follows. In \S\ref{sec: inflation} we outline the inflationary models considered in this work, before specifying our pipeline for analyzing them in \S\ref{sec: analysis}. Observational constraints are given in \S\ref{sec: results} before we conclude in \S\ref{sec: conclusion}. Throughout this work, we assume the \textit{Planck} 2018 cosmological parameters \citep{2020A&A...641A...6P}, with $h = 0.6732, \omega_b = 0.02238, \omega_c = 0.1201, \tau_{\rm reio} = 0.05431, n_s = 0.9660, A_s =2.101\times 10^{-9}$ and a single massive neutrino with $m_\nu = 0.06\,\mathrm{eV}$,. All inflationary predictions are computed in the scale-invariant limit, \textit{i.e.} with $n_s=1$, though the CMB estimators allow for non-zero $n_s-1$ when converting from dimensionless shape to curvature bispectrum.

\section{Inflationary Models}\label{sec: inflation}
\subsection{Strongly Coupled Sectors \& Unparticles}
\noindent Strongly coupled theories come in a variety of forms. One coarse way to parameterize theory space is to contrast the mass of the lowest excitation---the mass gap of the theory---to the Hubble scale during inflation. 
Then, we have three different scenarios: $\text{M}_{\text{Gap}}\ll H_{\text{inf}}$, $\text{M}_{\text{Gap}}\sim H_{\text{inf}}$ and $\text{M}_{\text{Gap}}\gg H_{\text{inf}}$. 
Since we already know that large anomalous dimensions are typically due to strong coupling, signals from large anomalous dimensions will become one of the most distinct signatures of the existence of strongly coupled dynamics. 
We focus on the $\text{M}_{\text{Gap}}\ll H_{\text{inf}}$ scenario, where the spectator sector can be well approximated by a CFT. Such CFT sectors may contain operators with large anomalous dimensions. There is no generic Lagrangian description for CFTs, and usually physicists work with conformal data, in the form of scaling dimensions and three-point coupling constants. For our purposes, specifying the conformal dimension will be enough, because we exchange the two-point function of the gapless sector, which is completely fixed by its scaling dimension. In other words, the computation and analysis we have is available for a general class of theories.

The scalar two-point function of an unparticle in flat space in Euclidean signature is completely fixed by conformal symmetries; it is given by 
\begin{equation}
    \langle \mathcal{O}_\Delta (\tau_1,\vec{x}_1)\mathcal{O}_\Delta (\tau_2,\vec{x}_2) \rangle_{\text{flat}} = \frac{1}{\Big((\tau_1 - \tau_2)^2 + (\vec{x}_1-\vec{x}_2)^2\Big)^\Delta}. 
\end{equation}
Here $\Delta$ is the \textit{scaling dimension} of the operator $\mathcal{O}_\Delta$, which relates to the difference between canonical scaling dimension of operator $\mathcal{O}$ and anomalous dimension that $\mathcal{O}$ picks up during the renormalization. 
Noting that the flat slicing of the de Sitter (dS) metric is conformally flat:
\begin{align}
    {\rm d} s^2 = \frac{-{\rm d} \eta^2 + {\rm d}\vec{x}^2}{H^2\eta^2} \equiv \alpha^2(\eta) (-{\rm d}\eta^2 + {\rm d}\vec{x}^2),
\end{align}
we can derive the two-point function of conformal fields in dS spacetime (as a good approximation for the inflationary two-point function) from the flat space result by taking $\tau\mapsto i\eta$ and the corresponding transformation of operators 
$\mathcal{O}_\Delta(x) \mapsto \tilde{\mathcal{O}}_\Delta(\alpha(\eta) x)\equiv\alpha(\eta)^{-\Delta} \mathcal{O}_\Delta$: 
\begin{align}
    \langle \tilde{\mathcal{O}}_\Delta (\eta_1,\vec{x}) \tilde{\mathcal{O}}_\Delta (\eta_2,\vec{y}) \rangle_{\text{dS}_4} &= \frac{\alpha(\eta_1)^{-\Delta}\alpha(\eta_2)^{-\Delta}}{\Big(-(\eta_1 - \eta_2)^2 + (\vec{x}-\vec{y})^2\Big)^\Delta} = \frac{(H^2\eta_1\eta_2)^\Delta}{\Big(-(\eta_1 - \eta_2)^2 + (\vec{x}-\vec{y})^2\Big)^\Delta}. 
\end{align}
In four dimensions, the unitarity bound of scalar conformal fields requires $\Delta\geq1$. 
For a specific model as proposed in \cite{Yang:2025apy}, the unparticle comes from a non-Abelian gauge theory in the dark sector. The theory lies on the edge of the conformal window, which becomes near-conformal and strongly coupled during inflation. 

Using the language of the EFT of inflation, the leading order interaction between the scalar perturbation $\pi$ and the scalar primary operator $\mathcal{O}_\Delta$ is 
\begin{align}
    &\mathcal{L}_{\pi\mathcal{O}} = \frac{1}{2}\mu^{2-\Delta} M_{\text{pl}}|{\dot{H}}|^{1/2} \left(-2\dot{\pi}\mathcal{O}_\Delta + (\partial_\mu\pi)^2\mathcal{O}_\Delta\right),
\end{align}
where $\mu$ is the effective coupling constant which will be absorbed in the overall size of the signal, i.e., $f_{\rm NL}^\Delta$. 
The quadratic interaction is what we start with to compute the correlator, as shown in \cite{Pimentel:2025rds}. 

\subsection{Bootstrap Predictions}
\noindent We begin by constructing a family of primordial bispectrum shapes arising from the tree-level exchange of a scalar unparticle with conformal dimension $\Delta\geq 1$. As discussed in \citep{Arkani-Hamed:2018kmz}, the induced curvature bispectrum can be obtained using bootstrap techniques, starting from the four-point function of a conformally coupled scalar $\varphi$. Writing the soft-limit of the dimensionless $\varphi$ four-point function as $b(u)$ (often denoted $F(u,1$)), the curvature three-point function is proportional to:
\beq
    B_\zeta(k_1,k_2,k_3) &\propto&  \frac{1}{k_1^3k_2^3}\left(1-\frac{k_1k_2}{k_{12}}\partial_{k_{12}}\right)\left(\frac{1-u^2}{u^2}\partial_ub_\Delta(u)\right) + \text{2 perms.}
\eeq
defining $u = k_3/k_{12}$ and $k_{12}\equiv k_1+k_2$. Following \citep{Pimentel:2025rds}, the four-point function of $\varphi$ for scalar unparticles can be computed analytically, leading to the following expression for $b(u)$:
\begin{align}
  b_\Delta(u)
  &\equiv
  -\frac{1}{\Delta - 1}
  \left( \frac{2u}{1+u} \right)^{2(\Delta-1)}
  {}_2F_1\!\left(1,\Delta-1;\Delta;\frac{2u}{1+u}\right)
  \nonumber\\
  &\quad
  +\frac{1}{2-\Delta}
  \left( \frac{2u}{1+u} \right)
  {}_2F_1\!\left(1,2-\Delta;3-\Delta;\frac{2u}{1+u}\right)
  \nonumber\\
  &\quad
  +\frac{2}{\Delta-1}
  \left( \frac{2u}{1+u} \right)
  {}_2F_1\!\left(1,1;\Delta;\frac{1-u}{1+u}\right),
\end{align}
where ${}_2F_1$ is the Gauss hypergeometric function. Using the curvature bispectrum, we can define the amplitude $f_{\rm NL}^{\Delta}$ and the dimensionless shape $S_\Delta(x,y)$:\footnote{We will denote the shape by both $S(kx,ky,k)$ and $S(x,y)$ in this work, invoking scale-invariance.}
\begin{align}
    B_{\zeta,\Delta}(k_1,k_2,k_3) \equiv \frac{18}{5}\frac{A_\zeta^2}{k_1^2k_2^2k_3^2}\,\times\,f_{\rm NL}^{\Delta}\,\times\,S_\Delta\left(\frac{k_1}{k_3},\frac{k_2}{k_3}\right)
\end{align}
normalizing in the equilateral limit, with $S_\Delta(1,1)=1$. Note that we assume the de-Sitter limit ($n_s=1$), with curvature power spectrum $P_\zeta(k)=A_\zeta/k^3$. In the squeezed limit, $S_\Delta$ has the scaling
\begin{align}
    \label{eq: scaling}
    S_\Delta(x,1) \sim \begin{cases} x^{\Delta-1} & \text{for } 1\leq \Delta\leq 2 \\
    x & \text{else.}\end{cases};
\end{align}
the latter matches the scaling of the single-field self-interaction templates, whilst $\Delta\to 1$ approaches the limiting form of the canonical orthogonal shape, as well as various non-standard vacuum scenarios (albeit transiently \cite{Holman:2007na,Meerburg:2009fi}).

Using \textsc{Mathematica}, we evaluate $S_\Delta(x,y)$ over the triangular domain $|x-y|\leq 1\leq x+y$ and $x\leq y\leq 1$, and tabulate the result on a logarithmic grid in $x\in[10^{-4},1]$ and a linear grid in $y\in[0.5,1]$ for a range of conformal dimensions $\Delta \in [1,\,9]$ (omitting integer $\Delta$, where $b_\Delta(u)$ diverges). These tabulated shapes form a library of bootstrap unparticle bispectra suitable for use in the analysis below.

\subsection{Phenomenology}
Whilst the above bispectra are sourced by a very different mechanism to those of mundane single-field self-interactions, it is \textit{a priori} unclear if they can be distinguished, particularly given the scaling-behavior of \eqref{eq: scaling}. To ascertain this, we can compute the primordial cosine between the unparticle shape functions, $S_\Delta$, and standard local, equilateral, and orthogonal shapes, defined via \cite{Gangui:1993tt,Wang:1999vf,Verde:1999ij,Komatsu:2001rj,Creminelli:2003iq,Zaldarriaga:2003my,Babich:2004gb}
\beq\label{eq: simple-templates}
    S_{\rm loc}(k_1,k_2,k_3) &=& \frac{1}{3}\left(\frac{k_1^2}{k_2k_3}+\text{2 perms.}\right)\\\nonumber
    S_{\rm eq}(k_1,k_2,k_3) &=& \left(\frac{k_1}{k_2}+\text{5 perms.}\right)-\left(\frac{k_1^2}{k_2k_3}+\text{2 perms.}\right)-2\\\nonumber
    S_{\rm orth}(k_1,k_2,k_3) &=& 3\left(\frac{k_1}{k_2}+\text{5 perms.}\right)-3\left(\frac{k_1^2}{k_2k_3}+\text{2 perms.}\right)-8.
\eeq
In this work, we will additionally consider the alternative orthogonal shape proposed in \citep{Senatore:2009gt} (which we label $\mathrm{orthogonal}^*$):
\beq\label{eq: orth*-shape}
    S_{\rm orth^*}(k_1,k_2,k_3) &=& (1+p)\frac{\tilde{\Delta}}{e_3}-p\frac{\Gamma^3}{e_3^2}
\eeq
defining 
\beq
    &&p = \frac{27}{\frac{743}{7(20\pi^2-193)}-21}, \qquad \tilde{\Delta}=(k_1+k_2-k_3)(k_2+k_3-k_1)(k_3+k_1-k_2)\\\nonumber
    && e_3=k_1k_2k_3, \qquad \Gamma = \tfrac{2}{3}(k_1k_2+k_2k_3+k_3k_1)-\tfrac{1}{3}(k_1^2+k_2^2+k_3^2).
\eeq
In the squeezed limit ($x\ll 1$), these have asymptotic scalings proportional to $x^{-1}, x^1, x^0$ and $x^1$ respectively.

As shown in \citep{Senatore:2009gt}, the EFT of inflation predicts two dominant self-interaction templates, induced by the $\dot{\pi}^3$ and $(\partial_i\pi)^2\dot{\pi}$ vertices in the primordial Lagrangian. Whilst these are not directly factorizable, they can be extremely well described by a linear combination of $S_{\rm eq}$ and $S_{\rm orth^*}$, given their asymptotic limits $\lim_{x\ll 1}S^{\dot{\pi}^3(x,1)}\sim \lim_{x\ll1}S^{(\partial_i\pi)^2\dot{\pi}} \sim x$. For this reason, LSS studies, which are particularly sensitive to squeezed limits through scale-dependent bias, have primarily adopted the $S_{\rm orth^*}$ template \citep[e.g.,][]{Cabass:2022wjy}. CMB analyses, on the other hand, typically use the $S_{\rm orth}$ template, which, in combination with $S_{\rm eq}$ still provides a fairly accurate basis for self-interactions. Since we will sometimes marginalize over self-interactions in this work (which is particularly sensitive to the template definition if the correlation coefficient is large), we principally adopt the LSS definition, $S_{\rm orth^*}$ in this work.

The overlap between templates is defined via the cosine
\beq\label{eq: inner-product}
    \mathrm{cos}(S,S') &\equiv& \frac{\av{S|S'}}{\sqrt{\av{S|S}\av{S'|S'}}}, \quad \av{S|S'} \equiv \int dk_1\int dk_2\int_{|k_1-k_2|}^{k_1+k_2}dk_3\,\frac{W(k_1)W(k_2)W(k_3)}{k_1+k_2+k_3}S(k_1,k_2,k_3)S'(k_1,k_2,k_3) 
\eeq
integrating over a triangular domain. To roughly mimic the sensitivity of \textit{Planck}, we add a factor of $(k_1+k_2+k_3)^{-1}$ to convert from three- to two-dimensions (following \citep{2009PhRvD..80d3510F}), and a set of weighting functions $W(k)$ to restrict to the relevant Fourier-space scales (with support for $2\times 10^{-4}\leq k/\mathrm{Mpc}^{-1}\leq 2\times 10^{-1}$), following \citep{Philcox:2025bbo}.

\begin{figure}
    \centering
    \includegraphics[width=0.8\linewidth]{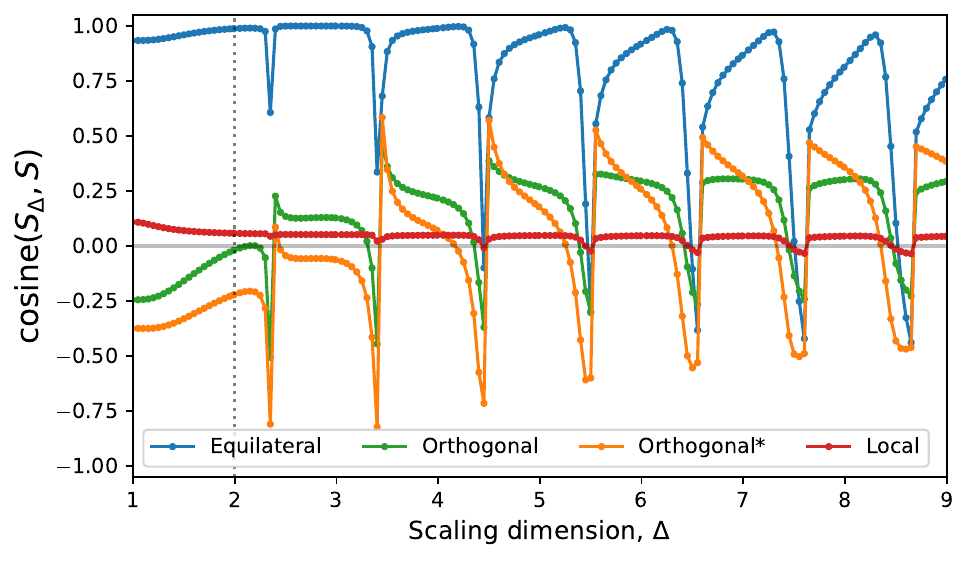}
    \caption{\textbf{Overlap of unparticle shapes with standard templates}. We plot three-dimensional cosine between the unparticle shape, $S_\Delta$, and various standard bispectrum shapes as a function of the conformal scaling dimension, $\Delta$. These cosines are obtained via a suitable inner-product in primordial-space, applying a mild $k$-weighting to mimic the \textit{Planck} sensitivity. The `orthogonal*' template refers to the shape of \citep{Senatore:2009gt} (used in most large-scale structure studies), which has the same squeezed-limit scaling as the equilateral shape. Notably, models with $\Delta<2$ exhibit an enhanced squeezed limit, though unitarity requires a lower bound of $\Delta\geq 1$.}
    \label{fig: cosines}
\end{figure}

In Fig.\,\ref{fig: cosines}, we plot the cosines for a range of scaling dimensions. Whilst the correlations with the local shape are always weak ($\lesssim 10\%$), those with the equilateral template can be large, reaching $99.99\%$ for $\Delta\approx 3$.
The overlap with the two orthogonal shapes is smaller than for equilateral, but generically increases with $\Delta$, whilst that with $S_{\rm eq}$ decreases. We also find larger correlations as we approach $\Delta = 1$, which is attributed to the enhanced squeezed limit (which differs considerably from the equilateral shape). Notable features are observed around half-integer values of $\Delta$, where the cosine with the equilateral shape drops considerably and that with the orthogonal shapes is enhanced. This matches the behavior found in \citep{Pimentel:2025rds}, and corresponds to the emergence of oscillations in the shapes (outside the squeezed regime, and thus distinct from the weakly-coupled cosmological collider signatures); these can cause the equilateral limit, $S(1,1)$, to vanish, which we will find to source strong $f_{\rm NL}$ bounds in the sections below. Overall, it is clear that, whilst the unparticle templates overlap strongly with the equilateral shape at low-$\Delta$, they present novel phenomenology at both specific values of $\Delta$ (around half-integers) and for $\Delta\gg2$, which are worthy of further study.

\section{Analysis Pipeline}\label{sec: analysis}
\subsection{Orthogonalization}\label{subsec: orthog}
\noindent How can we efficiently search for the signatures of unparticles across a wide range of scaling dimensions? From Fig.\,\ref{fig: cosines}, it is clear that to fully probe the unparticle regime, we require both a wide and dense sampling in $\Delta$. Directly analyzing each model in question is inefficient, particularly given the high-correlations between templates, thus we advocate performing a basis decomposition in order to isolate the novel unparticle signatures using a small set of templates.

We represent a given unparticle template $S_\Delta$ as a linear combination of the equilateral and orthogonal$^*$ templates, plus a sum of residual components expanded in a carefully-chosen basis (obtained via a singular value decomposition (SVD), as specified below):
\beq\label{eq: basis-decomposition}
    S_{\Delta}(k_1,k_2,k_3) &\approx & \sum_{i=1}^{N_{\rm SVD}+2}f_i(\Delta) S_{i}(k_1,k_2,k_3), \qquad \{S_i\} = \{S_{\rm eq}, S_{\rm orth^*}, S^{(1)}_{\rm SVD}, \cdots, S_{\rm SVD}^{(N_{\rm SVD})}\}.
\eeq
Here, the basis elements $S_{i}(k_1,k_2,k_3)$ are independent of $\Delta$, and the $k$-independent weights solve
\beq
    \av{S_\Delta|S_j} \equiv \sum_{i=1}^{N_{\rm SVD}+2}f_i(\Delta)\av{S_i|S_j} \qquad j=1,2,\cdots,N_{\rm SVD}+2
\eeq
using the inner product of \eqref{eq: inner-product}. Since we explicitly include $S_{\rm eq}$ and $S_{\rm orth^*}$ in the basis set (similar to \citep{Suman:2025tpv,Suman:2025vuf}), the decomposition of \eqref{eq: basis-decomposition} fully captures the correlations of unparticle models with single-field templates, with $\{S_{\rm SVD}^{(i)}\}$ encoding the novel unparticle signatures.\footnote{Note that this would not hold if we had used the $S_{\rm orth}$ template instead of $S_{\rm orth^*}$. Whilst the distinction is small in practice, it is important when one wishes to marginalize over self-interactions.} 

To implement the above prescription in practice, we first generate a grid of unparticle models across $N_\Delta=161$ values of the scaling dimension and $\mathcal{O}(10^6)$ triplets of wavenumbers. Using these, we define two $N_\Delta\times N_\Delta$ Fisher matrices:
\beq
    F^{\rm unmarg}_{\Delta\Delta'} &=& \av{S_\Delta|S_{\Delta'}}\\\nonumber
    F^{\rm marg}_{\Delta\Delta'} &=& \av{S_\Delta|S_{\Delta'}} - \begin{pmatrix}\av{S_\Delta|S_{\rm eq}} & \av{S_\Delta|S_{\rm orth^*}}\end{pmatrix}\begin{pmatrix}\av{S_{\rm eq}|S_{\rm eq}} & \av{S_{\rm orth^*}|S_{\rm eq}} \\ \av{S_{\rm orth^*}|S_{\rm eq}} & \av{S_{\rm orth^*} | S_{\rm orth^*}}\end{pmatrix}^{-1}\begin{pmatrix}\av{S_{\rm eq}|S_\Delta'} \\ \av{S_{\rm orth^*}|S_\Delta'}\end{pmatrix};
\eeq
these are proportional to the (idealized) inverse-square error bar on $f_{\rm NL}^{\Delta}$ before and after marginalizing over $f_{\rm NL}^{\rm eq}$ and $f_{\rm NL}^{\rm orth^*}$ (\textit{i.e.}\ inflationary self-interactions). 
Next, we perform an eigendecomposition of $F_{\Delta\Delta'}^{\rm marg}$. Restricting to the first few eigenmodes defines our SVD basis: $\{S_{\rm SVD}^{(i)}\}$.\footnote{An alternative approach would be to construct a basis from $F^{\rm unmarg}$ directly. This is less efficient, however, and carries no guarantee of preserving information after marginalizing over self-interactions.} By construction this basis is orthogonal to $S_{\rm eq}$ and $S_{\rm orth^*}$ (and is itself orthonormal), implying that it captures only the independent unparticle signatures, as desired.\footnote{This basis can be equivalently derived by performing a projecting out $S_{\rm eq}$ and $S_{\rm orth^*}$ from $\{S_\Delta\}$ using a Gram-Schmidt procedure, then performing a singular value decomposition on the deprojected shapes.} 

Given the distinct squeezed limits of \eqref{eq: scaling}, it is advantageous to treat the $1\leq \Delta<2$ and $\Delta\geq2$ separately, constructing a separate SVD basis for each. For the former, we also include $S_{\rm orth}$ in the basis set (which has the same squeezed limit as $S_{\Delta=1}$) which reduces the number of SVD templates needed to accurately reconstruct the shape. Here, we require that the SVD basis reproduces both $F^{\rm unmarg}$ and $F^{\rm marg}$ to within $98\%$: this requires $N_{\rm SVD}=2$ for $1\leq \Delta<2$ and $N_{\rm SVD}=5$ for $\Delta\geq 2$. 

\begin{figure}
    \centering
    \includegraphics[width=0.7\linewidth]{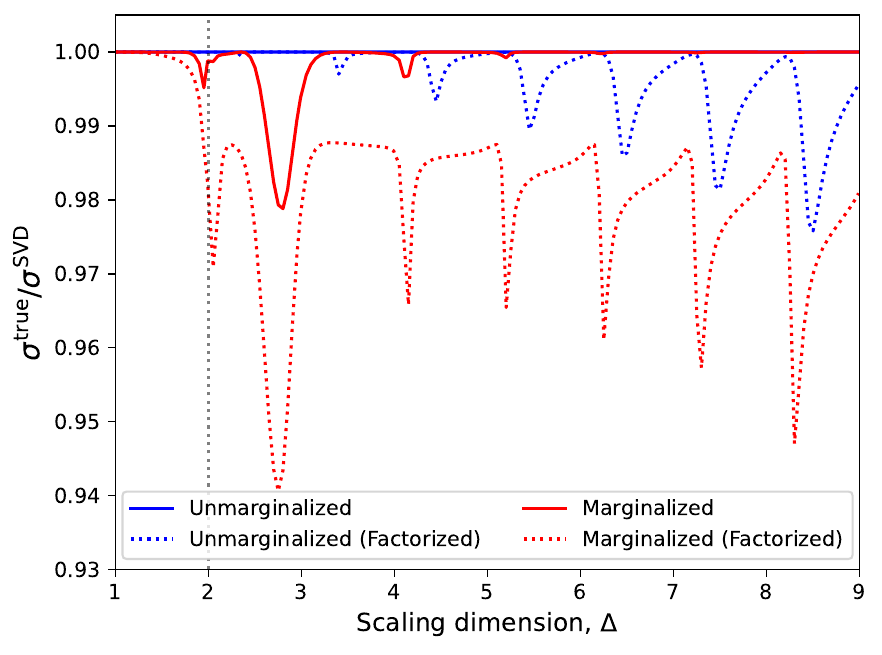}
    \caption{\textbf{Information loss from basis projection}. We plot the theoretical errorbars on the unparticle shape from an idealized forecast ($\sigma^{\rm true}$) and after projecting onto the orthogonalized basis ($\sigma^{\rm SVD}$). The blue lines show the results for a single-template analysis of $f_{\rm NL}^{\Delta}$, whilst the red marginalize over the two self-interaction shapes (via $f_{\rm NL}^{\rm eq}$ and $f_{\rm NL}^{\rm orth^*}$). The dotted lines show the errorbars after projecting the models onto the neural-factorized basis -- the imperfect reconstruction results in a slight loss of information for some choices of $\Delta$. In all cases, we compute the errorbar using a three-dimensional Fisher forecast, which allows us to directly analyze both factorized and unfactorized templates.}
    \label{fig: information-loss}
\end{figure}

In Fig.\,\ref{fig: information-loss} (solid lines), we plot the information loss incurred by our compression onto the SVD basis, measured via the idealized errorbars $\sigma(f_{\rm NL}^{\Delta}) \propto F_{\Delta\Delta}^{-1/2}$. For the full templates, we find essentially no loss of information (below one part in $10^5$), whilst the marginalized constraints are preserved to within $2\%$. The largest deviations are found for $\Delta\to 2$ (where the low-$\Delta$ basis is least efficient) or for $\Delta\sim 2.5$ (where the non-self-interaction behavior is most apparent). All in all, we find that we can decompose the full set of $161$ models to just $7$ templates without negligible loss of information, even after marginalizing over self-interactions.

\subsection{Factorization}
\noindent We now consider how to practically constrain the basis shapes derived above using temperature and polarization anisotropy data. Due to the two-dimensional nature of the CMB and the underlying oscillatory projection integrals, efficient data analyses are possible only if the shape in question is factorizable, meaning that it can be expanded in the (permutation-invariant) form
\beq\label{eq: factorizable-shape}
    S_{\rm fact}(k_1,k_2,k_3) &=& \frac{1}{6}\sum_{n=1}^{N_{\rm fact}}w_n\alpha_n(k_1)\beta_n(k_2)\gamma_n(k_3) + \text{5 perms.},
\eeq
where $\alpha_n,\beta_n,\gamma_n$ are arbitrary functions and $w_n$ are scalar weights. Whilst the simplest templates (including those of \eqref{eq: simple-templates}) can be expressed in this form, most realistic models cannot. To proceed, one must rewrite the desired model in a separable form, for example, expanding it in a modal basis \citep[e.g.,][]{2009PhRvD..80d3510F}. In this work, we adopt the approach of \citep{Philcox:2025bbo}, which uses machine learning tools to construct a low-dimensional set of neural network basis functions, $\{\alpha_n,\beta_n,\gamma_n\}$, and weights, $\{w_n\}$, tailored to the target bispectrum. This has been shown to accurately reproduce a variety of bispectra using just a handful of terms (typically $N_{\rm fact}\lesssim 5$), which is far fewer than for the modal approach \citep[e.g.,][]{Sohn:2024xzd}. Here, we apply these techniques to each of the SVD basis functions individually: an alternative approach would be to construct a joint basis directly from the $S_\Delta$ models, though this is more expensive to optimize.

We implement the factorization procedure using the \textsc{separable\_bk} code developed in \citep{Philcox:2025bbo}.\footnote{Available at \href{https://github.com/kunhaozhong/separable_bk}{github.com/kunhaozhong/separable\_bk}.} For each of the SVD templates described above, we optimize for the basis functions using stochastic gradient descent given the loss function:
\beq\label{eq: ML-loss}
    \mathcal{L}({\rm basis}) &=& \frac{\av{S_{\rm fact}[{\rm basis}]-S_{\rm SVD}|S_{\rm fact}[{\rm basis}]-S_{\rm SVD}}}{\av{S_{\rm SVD}|S_{\rm SVD}}}
\eeq
where $\av{\cdot|\cdot}$ is the inner product defined in \eqref{eq: inner-product}, and `basis' contains the set of $3N_{\rm fact}$ one-dimensional functions used in \eqref{eq: factorizable-shape}, which are parametrized as neural networks with a single hidden layer. We introduce a number of improvements compared to the former work. Firstly, instead of learning the weights $\{w_n\}$ dynamically alongside $\alpha_n,\beta_n,\gamma_n$, we fix them to the maximum-likelihood solution (obtained by setting $\delta\mathcal{L}/\delta w_n = 0$). The optimal weights are thus
\beq
    \av{M_n|S_{\rm SVD}}=\sum_{n'=1}^{N_{\rm fact}}w^{\rm opt}_{n'}\av{M_n|M_{n'}},\qquad M_n(k_1,k_2,k_3) \equiv \alpha_n(k_1)\beta_n(k_2)\gamma_n(k_3)+\text{5 perms.}
\eeq
This leads to faster training and a more accurate model. Secondly, we add a number of penalties to the loss function of \eqref{eq: ML-loss}:
\beq
    \mathcal{L}(\rm basis) &\supset& 10^{-3}\log\left|\sum_{n=1}^{N_{\rm basis}}\mathcal{G}_{nn}^2\right| + 10^{-1}\frac{\sum_{n,n'=1}^{N_{\rm basis}}\left(\G_{nn'}-\delta^{\rm K}_{nn'}\G_{nn}\right)^2}{\sum_{n=1}^{N_{\rm basis}}\G_{nn}^2} + 10^{-2}\left(\log\det \G-1\right)^2
\eeq
for $\G_{nn'}\equiv \av{M_n|M_{n'}}$. Respectively, these ensure that the overall amplitudes of the $N_{\rm fact}$ basis functions are close to unity (noting that scaling degeneracy with $w_n$), that the basis is approximately orthogonal (with $\G_{nn'}\to 0$ for $n\neq n'$), and that degenerate modes are avoided. Whilst this correction is certainly \textit{ad hoc}, we have found it to substantially improve the training efficacy. 

We find that we can obtain $95\%$ accurate factorizable representations of the seven SVD shapes discussed above using $N_{\rm fact}=5$ terms. For each shape, computation required $\approx 2$ hours on a single CPU core, or $\sim 30$ minutes on a GPU. Our SVD templates proved significantly more difficult to analyze than the weakly-coupled collider shapes considered in the previous work \citep{Philcox:2025bbo}, which could be reproduced to $>99.9\%$ accuracy with a similar $N_{\rm fact}$. We attribute this to the complex phenomenology of numerical SVD-based templates and the oscillations in the unparticle shapes themselves,\footnote{In initial testing, we considered applying the factorization scheme to the unparticle shapes themselves, rather than the SVD basis. This yielded more accurate decompositions, but is expensive to apply in practice, due to the large number of scaling dimensions considered in this work.} though note that $95\%$ accuracy is more than sufficient in practice.

Finally, we can express the unparticle shapes discussed above in our orthogonalized-plus-factorized basis. This is practically achieved by replacing $S_{\rm SVD}^{(i)}$ in \eqref{eq: basis-decomposition} with the neural factorized representations obtained from \eqref{eq: factorizable-shape}. By computing the idealized Fisher matrix using the new basis functions, we can assess the information loss induced by the projections: this is shown in the dotted lines in Fig.\,\ref{fig: information-loss}. We find that the direct constraints on $f_{\rm NL}^{\Delta}$ are preserved to $98\%$ accuracy (with largest deviations found at large $\Delta$), whilst those marginalized over self-interactions are preserved to within $94\%$ (with greatest information loss found for $\Delta\sim 2.75$). All in all, we conclude that the basis transformations performed above (which are necessary to compare theory and data) do not severely impact our ability to constrain the unparticle phenomenology.

\subsection{Estimation}
\noindent The outcome of the previous sections is a set of seven SVD basis shapes, each of which has been decomposed into $N_{\rm fact}=5$ factorized components, allowing them to be readily constrained from observational data. To do so, we utilize the \textsc{PolySpec} code \citep{PolyBin,Philcox4pt2,Philcox:2023psd,Philcox:2023uwe} (building on \citep{2011MNRAS.417....2S,2015arXiv150200635S}), which implements optimal Komatsu-Spergel-Wandelt (KSW) \citep[e.g.,][]{Komatsu:2003iq} estimators for a variety of primordial bispectrum and trispectrum models. Here, we use the general factorized shape functionality discussed in \citep{Philcox:2025bbo}, which takes the basis functions and weights computed by the \textsc{separable\_bk} code as input and outputs the corresponding $f_{\rm NL}$ estimates.\footnote{Notably, \textsc{PolySpec} transforms from dimensionless to dimensionful primordial bispectrum ($S\to B_\zeta$) using $n_s\approx 0.96$, to ensure consistency with the power spectrum scalings.} Schematically, the estimator takes the form $\widehat{f}_{\rm NL} = \F^{-1}\widehat{N}$, defining the numerator and normalization
\beq\label{eq: estimator}
    \widehat{N}_n &\sim& \frac{1}{6}\sum_{ijk}\frac{\partial \av{a_ia_ja_k}}{f^{(n)}_{\rm NL}}\left([C^{-1}a]_i[C^{-1}a]_j[C^{-1}a]_k-3[C^{-1}a]_iC^{-1}_{jk}\right)\\\nonumber
    \F_{nn'} &\sim& \frac{1}{6}\sum_{ijki'j'k'}\frac{\partial \av{a_ia_ja_k}}{f_{\rm NL}^{(n)}}C^{-1}_{ii'}C^{-1}_{jj'}C^{-1}_{kk'}\frac{\partial \av{a_{i'}a_{j'}a_{k'}}}{f^{(n')}_{\rm NL}},
\eeq
for a template specified by the amplitude $f_{\rm NL}^{(n)}$, where $a_i$ is the CMB datavector, $C$ is its covariance, and $i$ indexes pixels (or harmonic coefficients) and polarizations. If the primordial bispectrum is factorizable (as in \eqref{eq: factorizable-shape}), the $i,j,k$ summations decouple, allowing the expression to be efficiently computed using spherical harmonic transforms. By construction the estimator is (almost) minimum-variance, implying that its normalization satisfies the Cram\'er-Rao bound: $\sigma(f_{\rm NL}^{(n)})\geq \F_{nn}^{-1/2}$.

To facilitate marginalization of self-interaction shapes, we add $f_{\rm NL}^{\rm orth}$ and $f_{\rm NL}^{\rm orth^*}$ estimators to \textsc{PolySpec}, building on the pre-existing $f_{\rm NL}^{\rm loc}$ and $f_{\rm NL}^{\rm eq}$ implementations. The former is a simple adjustment of the equilateral estimator (as is clear from the similarities in the templates of \eqref{eq: simple-templates}) and matches previous works \citep[e.g.,][]{Planck:2019kim}. To our knowledge, the $f_{\rm NL}^{\rm orth^*}$ estimator has not been previously implemented in CMB analyses owing to its more complex shape \eqref{eq: orth*-shape} \citep[cf.][]{Senatore:2009gt}, though remains possible since $S_{\rm orth^*}$ can be expressed as a sum of monomials and thus factorized explicitly:
\beq
    S_{\rm orth^*}(k_1,k_2,k_3) &=& \sum_{a+b+c=0}\omega_{abc}k_1^{a}k_2^{b}k_3^{c}
\eeq
where $-2\leq a,b,c\leq 4$. Due to the larger exponents in the expansion for $S_{\rm orth^*}$ compared to that of $S_{\rm eq}$ or $S_{\rm orth}$ (which have $-1\leq a,b,c\leq 2$), we find the $f_{\rm NL}^{\rm orth^*}$ analysis to be numerically sensitive to the large-$k$ contributions to the CMB transfer functions in highly squeezed regimes: to avoid this issue we restrict our attention to $\ell_{\rm min}=100$ in this work.

Our dataset is the \textit{Planck} PR4 temperature anisotropy and $E$-mode polarization measurements, as obtained from the \textsc{npipe} processing pipeline \citep{Planck:2020olo} (matching a number of recent non-Gaussianity studies \citep[e.g.,][]{Philcox4pt3,Philcox:2025bbo,Philcox:2025lxt,Jung:2025nss}). To avoid foreground contamination, we apply the common component separation mask in both temperature and polarization (inpainting small holes linearly), and work with component-separated maps obtained via the \textsc{sevem} prescription \citep{Planck:2018yye}. Data are filtered in harmonic-space using a translation-invariant power spectrum comprising the best-fit theory spectrum, the observed beam and transfer functions, and the noise (itself obtained from half-mission maps), and we restrict to $\ell\in[100,2048]$. To compute errorbars, we employ $100$ FFP10/\textsc{npipe} simulations \citep{Planck:2020olo,Planck:2015txa}, with a further $100$ used to compute the linear term in \eqref{eq: estimator}. 

Given the \textit{Planck} dataset and simulations, we jointly infer the $f_{\rm NL}$ amplitudes corresponding to the equilateral, orthogonal$^*$, and SVD templates, as well as the ISW-lensing amplitude, which is a known contaminant to primordial analyses \citep{Philcox:2025lxt}. We perform two sets of analyses: one for $\Delta<2$ (analyzing also the orthogonal template, as discussed above), and one for $\Delta\geq 2$. Since the linear term of the bispectrum estimator only impacts squeezed configurations and is expensive to compute (involving an average over many Monte Carlo simulations), we omit this term when analyzing the $\Delta\geq 2$ shapes, noting that the squeezed limit is highly suppressed (cf.\,\ref{eq: scaling}) and thus the cubic term in \eqref{eq: estimator} is already essentially optimal. Due to the various computational tricks employed within \textsc{PolySpec}, the full computation is fast, requiring around $20$ node-minutes per simulation (involving a few thousand spherical harmonic transforms), with an extra $5$ node-hours required to compute the mask-, noise- and beam-dependent factor $\F$. 

The final step is to transform back from the SVD basis to the unparticle amplitudes. This is simply achieved by recasting \eqref{eq: estimator} as an estimator for $f_{\rm NL}^{\Delta}$ by inserting the basis expansion \eqref{eq: basis-decomposition}:
\beq
    \widehat{f}^{\Delta,\,\rm unmarg}_{\rm NL} &=& \F_{\Delta\Delta}^{-1}\widehat{N}_\Delta, \qquad \widehat{N}_\Delta =\sum_{i=1}^{N_{\rm SVD}+2}f^{(i)}(\Delta)\widehat{N}_i, \qquad \F_{\Delta\Delta} = \sum_{i,j=1}^{N_{\rm SVD}+2}f^{(i)}(\Delta)f^{(j)}(\Delta)\F_{ij},
\eeq
where $\widehat{N}_i$ and $\F_{ij}$ (for $i=1,\cdots,N_{\rm SVD}+2$) are the estimator numerators and normalizations computed previously, and $f_i(\Delta)$ are the weights obtained in \S\ref{subsec: orthog} (computed using the factorized approximations to the SVD templates). Here, $\widehat{f}^{\Delta,\rm unmarg}_{\rm NL}$ is the estimated amplitude for the unparticle model with scaling dimension $\Delta$, whilst $\F_{\Delta\Delta}^{-1/2}$ is its (mask-, beam- and noise-dependent) optimal errorbar. Notably, we treat each value of $\Delta$ independently, \textit{i.e.}\ we perform single-template analyses, ignoring correlations between templates with different scaling dimensions.
By performing joint analyses of $f_{\rm NL}^{\Delta}$ with the equilateral and orthogonal$^*$ amplitudes, we can similarly compute constraints marginalized over self-interactions:
\beq
    \begin{pmatrix}
        \widehat{f}^{\rm eq,\,marg}_{\rm NL}\\
        \widehat{f}^{\rm orth^*,\,marg}_{\rm NL}\\
        \widehat{f}^{\Delta,\,\rm marg}_{\rm NL}\end{pmatrix}
        &=&
    \begin{pmatrix}
        \F_{11} & \F_{12} & \F_{1\Delta}\\
        \F_{21}  & \F_{22} & \F_{2\Delta} \\ \F_{\rm \Delta1} & \F_{\Delta2}  & \F_{\rm \Delta\Delta}
    \end{pmatrix}^{-1}
    \begin{pmatrix}
        \widehat{N}_1\\
        \widehat{N}_2\\
        \sum_{i=1}^{N_{\rm SVD}+2}f^{(i)}(\Delta)\widehat{N}_i\\
    \end{pmatrix}
\eeq
where $i=1,2$ correspond to equilateral and orthogonal$^*$ templates, and we have defined
\beq
    \F_{I\Delta} = \sum_{j=1}^{N_{\rm SVD}+2}f^{(j)}(\Delta)\F_{Ij}, \qquad \F_{\Delta J} = \sum_{i=1}^{N_{\rm SVD}+2}f^{(i)}(\Delta)\F_{iJ}.
\eeq
In this case, the optimal errorbar is given by $\left[\F^{-1}\right]^{1/2}_{\Delta\Delta}$, where $\F$ is the $3\times3$ matrix shown above. 

\section{Results}\label{sec: results}
\begin{figure}
    \centering
    \includegraphics[width=0.49\linewidth]{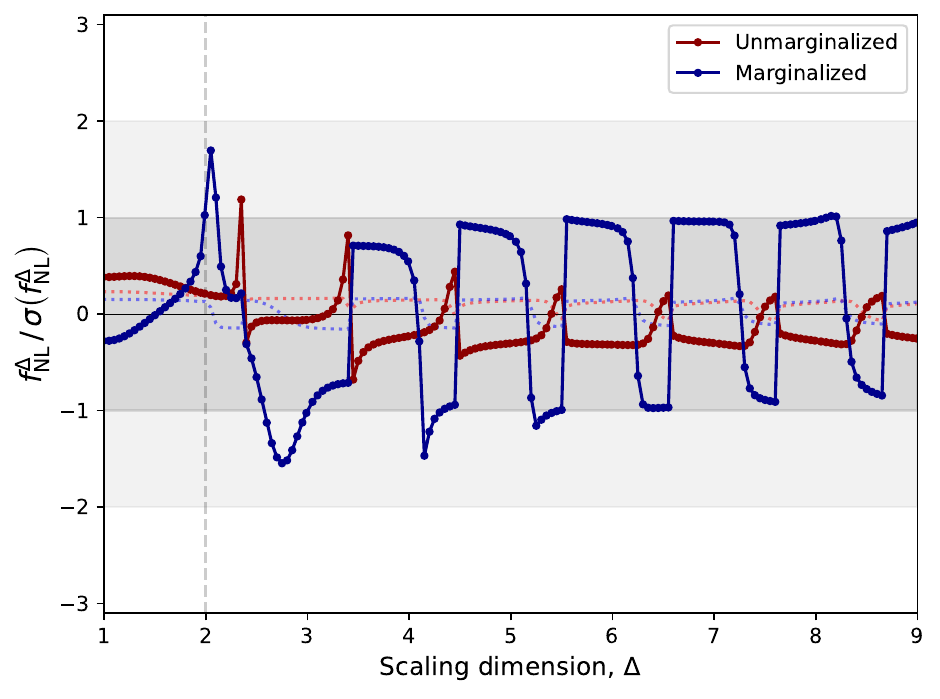}
    \includegraphics[width=0.49\linewidth]{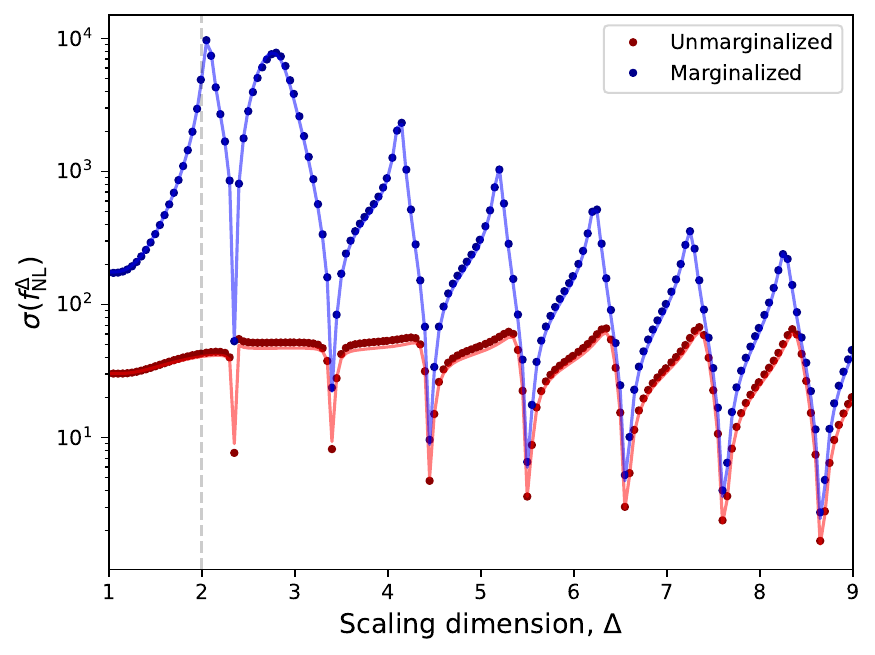}
    \caption{\textbf{CMB constraints on $f_{\rm NL}^{\Delta}$}. 
    \textit{Left panel}: Normalized constraints on $f_{\rm NL}^{\Delta}$ obtained from the \textit{Planck} PR4 temperature and E-mode polarization dataset. We show results for both the raw and self-interaction-marginalized datasets, which are obtained using the orthogonalized and factorized templates discussed in the main text. Dotted lines show the results obtained from the mean of $100$ FFP10/\textsc{npipe} simulations. Note that the constraints with $\Delta\leq 2$ and $\Delta>2$ use different basis sets for maximal efficiency. As shown in Fig.\,\ref{fig: corrmat}, the individual constraints are highly degenerate, and the small amplitudes are driven by the non-detection of $f_{\rm NL}^{\rm eq}$. We find no evidence for unparticles in \textit{Planck} data. 
    \textit{Right panel}: $1\sigma$ errors on the unparticle template amplitude. The points show results computed using 100 FFP10/\textsc{npipe} simulations, whilst the lines show the theoretical errors obtained from the (two-dimensional) Fisher matrix. The close agreement between points and lines indicate that our analysis is close-to-optimal. The constraints are strongest for certain values of $\Delta$ (around half-integers) where the equilateral cosine in Fig.\,\ref{fig: cosines} is small; these points also represent the most promising discovery space, given the limited correlations with self-interaction shapes.}
    \label{fig: errorbars}
\end{figure}

\noindent We now present results of the analysis described in \S\ref{sec: analysis}. The left panel of Fig.\,\ref{fig: errorbars} shows the constraints on the unparticle bispectrum amplitudes, $f_{\rm NL}^{\Delta}$, obtained using the pipeline of this work. Both before and after marginalization of self-interaction templates, we find no detection of unparticles, with a maximum signal-to-noise of $1.2\sigma$ at $\Delta=2.35$ (unmarginalized) or $1.7\sigma$ at $\Delta=2.05$ (marginalized). Notably, the latter occurs for an unparticle model which is highly degenerate with  self-interactions (as seen from the large $\sigma(f_{\rm NL}^\Delta)\sim 10^4$), thus it highly indicative of a noise fluctuation. When analyzing the mean of the mocks, we also find null detections (with a maximum significance of $0.23\sigma$ (unmarginalized) and $0.17\sigma$ (marginalized)), implying that our estimators do not induce additive bias. Despite the large number of models analyzed ($161$ in total), the maximum detection significance is weak: this occurs since the various models are highly correlated, and the unmarginalized constraints are largely informed by the $f_{\rm NL}^{\rm eq}$ posterior (cf.\,Fig.\,\ref{fig: cosines}), which is centered close to zero. We also find no detection of the self-interaction templates (analyzed independently): with our datacuts (including $\ell_{\rm min}=100$), we find $f_{\rm NL}^{\rm eq}=-16\pm51$ and $f_{\rm NL}^{\rm orth^*} = -12\pm12$ (and $f_{\rm NL}^{\rm orth} = -25\pm24$). Due to the negligible correlations between these two self-interaction templates, these constraints are unchanged if we analyze $f_{\rm NL}^{\rm eq}$ and $f_{\rm NL}^{\rm orth^*}$ jointly.

From the right panel of Fig.\,\ref{fig: errorbars} it is clear that the errorbar on $f_{\rm NL}^{\Delta}$ varies considerably with the scaling dimension, $\Delta$. Before marginalizing over self-interactions, we find $\sigma(f_{\rm NL}^{\Delta})\sim 30-60$ across most of the parameter space, with weaker bounds for large $\Delta$ and an enhancement as $\Delta\to 1$ (which has a stronger squeezed limit, \citep[cf.,][]{Kalaja:2020mkq}). For particular values of $\Delta$ (close to the half-integers, as in \citep{Pimentel:2025rds}), the errorbar is significantly tightened, reaching $\sigma(f_{\rm NL}^{\Delta}$$) = 1.7$ for $\Delta = 7.65$. In general, the quoted constraining power (but not the signal-to-noise) depends strongly on our choice of normalization: we find lowest $\sigma(f_{\rm NL}^{\Delta})$ for models with smallest equilateral bispectra (with $\sigma(f_{\rm NL}^{\Delta})\to 0$ if $S_\Delta(1,1)\to 0$). Notably, the errorbars obtained from simulations closely match those predicted from the estimator normalization, $\F_{\Delta\Delta}$ (within $12\%$ in all cases), indicating that our estimator is close to its Cram\'er-Rao bound, and thus (almost) optimal.

When marginalizing over the equilateral and orthogonal$^*$ templates, the constraints on $f_{\rm NL}^{\Delta}$ degrade by factors ranging from $1.3$ to $220$, depending on the scaling dimension, but remain essentially optimal. This highlights the strong degeneracies between the unparticle models and single-field self-interactions for certain values of $\Delta$ and matches the conclusions drawn from the idealized primordial cosines shown in Fig.\,\ref{fig: cosines} (and simple forecasts obtained from the corresponding Fisher matrices). For much of the parameter space, the CMB  bispectrum is unable to distinguish between unparticle exchange and single-field self-interactions. For values of $\Delta$ close to half-integers and $\Delta\gg 1$, this is not the case, and marginalization degrades the constraints only partially. For these models, the unparticle phenomenology is markedly different to that of known templates and our analysis is adding novel information. Such signatures represent important targets for future studies with higher-resolution datasets.

\begin{figure}
    \centering
    \includegraphics[width=0.85\linewidth]{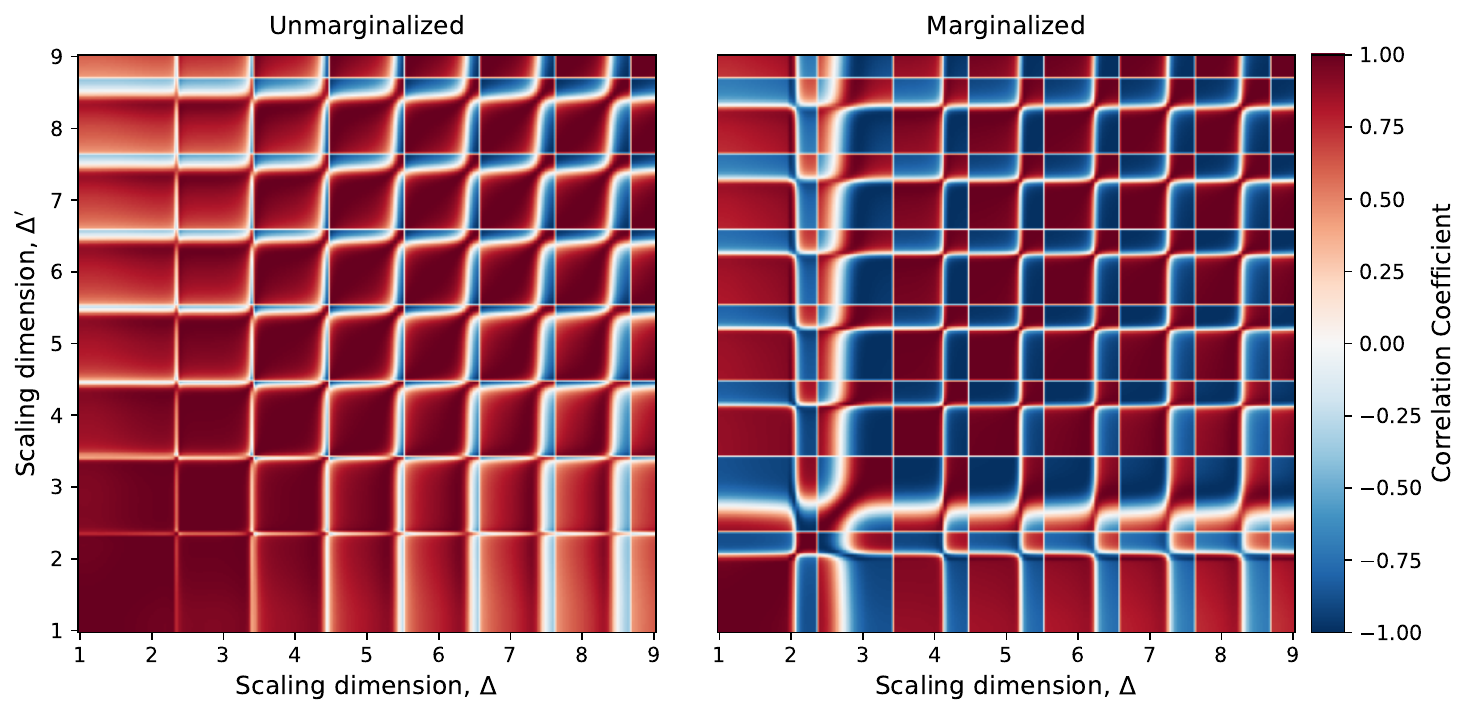}
    \caption{\textbf{Correlation matrix for $f_{\rm NL}^{\Delta}$}. We compute correlations using $100$ FFP10/\textsc{npipe} simulations, both analyzing each template independently (left panel) and marginalizing over $f_{\rm NL}^{\rm eq}$ and $f_{\rm NL}^{\rm orth^*}$. Correlations between adjacent points are very large, though we find minimal correlations for half-integer $\Delta$, indicating that the unparticle phenomenology in these regimes is particularly novel.}
    \label{fig: corrmat}
\end{figure}

Finally, we consider the correlation between the various $f_{\rm NL}^{\Delta}$ measurements. As shown in Fig.\,\ref{fig: corrmat}, the unmarginalized estimates for different scaling dimensions are highly covariant, particularly at low $\Delta$ (with, for example, $>95\%$ correlations between all models with $\Delta\leq 2.35$). Much of this is driven by the strong overlap of the templates with $f_{\rm NL}^{\rm eq}$ (Fig.\,\ref{fig: cosines}), which dominates for small $\Delta$. Consistent with the picture developed above, certain values of $\Delta$ exhibit minimal correlations with other models, with the size of the uncorrelated region increasing at larger $\Delta$. We find similar results for the marginalized constraints: the correlations between templates are large across most of the parameter space, including for small $\Delta$ (which has essentially just one independent component), but vanish for certain scaling dimensions, such as $\Delta\approx 2.5$. This indicates that the unparticle sector contains multiple distinguishable phenomenologies, suggesting that the value of $\Delta$ could, potentially, be probed using future high-resolution experiments.

\section{Discussion}\label{sec: conclusion}
\noindent In this work, we have performed the first direct searches for strongly-coupled sectors in inflation. Focusing on the unparticle scenario, which mixes a gapless conformal field theory with the inflaton, we have carefully examined the phenomenology of the exchange bispectra for both the primordial curvature perturbation and the CMB temperature and polarization anisotropies, utilizing bootstrap techniques developed in \citep{Pimentel:2025rds}. For many values of the scaling dimension $\Delta\geq 1$, the induced three-point functions are practically indistinguishable from those single-field self-interactions; however, for certain values of $\Delta$, novel phenomenology arises (as found in \cite{Pimentel:2025rds}). For $\Delta\leq 2$, we find enhanced squeezed limits that break the Maldacena consistency relation, mimicking weakly-coupled scalars with mass $m\leq \sqrt{2}H$. Moreover, for $\Delta$ close to half-integers and for $\Delta\gg1$, we find large oscillations, which are almost orthogonal to the single-field scenario (as well as those from massive field exchange). While the precise shapes of the correlators will depend on the underlying inflationary model, the overall conclusion is clear: strongly-coupled sectors can provide new, and potentially detectable, signals in the cosmological collider.

Much of this work has been devoted to constructing efficient pipelines for analyzing classes of inflationary models. Here, we advocate for a multi-step approach to obtain close-to-optimal $f_{\rm NL}$ constraints: (1) define a set of primordial bispectra of interest (computed using analytical or numerical techniques, across the full two-dimensional shape-space); (2) decompose these into a small set of basis templates using a PCA; (3) factorize the resulting basis templates using the neural-network-based approach of \citep{Philcox:2025bbo} (see \cite{Sohn:2023fte} for an alternative approach); (4) insert them into a KSW-style CMB bispectrum estimator, such as \textsc{PolySpec}. For the unparticle models, we have demonstrated that this pipeline can facilitate quasi-optimal analysis of a large set of models with negligible loss of signal-to-noise; indeed, we were able to search for 161 scaling dimensions using just 7 factorizable templates, preserving information to within $94\%$. A crucial part of our approach was to include the self-interaction shapes within the basis set (parametrized by $f_{\rm NL}^{\rm eq}$ and $f_{\rm NL}^{\rm orth}$, using the \cite{Senatore:2009gt} definition for the latter): this allows us to exactly account for the (often very large) cosines, avoiding possible numerical inaccuracies.\footnote{A similar approach was used in \citep{Suman:2025tpv,Suman:2025vuf} to isolate the novel features of weakly-coupled massive particle exchange by building orthogonalized templates. In this work, however, we opt to marginalize over self-interactions, which allows us to easily assess the model degeneracies.} Notably, we found that obtaining factorized bispectrum templates (which are imperative for CMB studies) was more difficult than \citep{Philcox:2025bbo} suggested due to the numerically-derived basis vectors (which are significantly less smooth than most theoretical models); however, we were still able to obtain $>95\%$ accuracy reconstructions for all templates considered, following some small optimizations to the code (including explicit regularization terms). 

Combining the bootstrapped unparticle templates with the optimal CMB estimators and the latest \textit{Planck} dataset, we obtained tight constraints on the unparticle-exchange bispectra for a wide range of scaling dimensions (as well the equilateral and orthogonal$^*$ shapes, with the latter analyzed for the first time). Across $1\leq \Delta\leq 9$, we obtained a maximum detection significance of $1.2\sigma$ (unadjusted by look-elsewhere effects), which (as usual) indicates no evidence for physics beyond the standard model. These constraints can be recast as bounds on the inflationary model, restricting the coupling of the inflaton to unparticles for a range of scaling dimensions. For the explicit ``Dark Walker'' scenario proposed in \cite{Yang:2025apy}, however, we find that our constraints are not competitive, since they do not improve on the upper limits from the model itself.
Due to the tight correlations between templates (which can reach $99.99\%$ in the CMB), many of the $f_{\rm NL}(\Delta)$ constraints are set by the \textit{Planck} equilateral bounds; however, we find tighter bounds for both low and (close-to) half-integer $\Delta$ matching the above discussion (and the analytic scalings of \citep{Kalaja:2020mkq}). Importantly, the latter constraints were degraded by only $\lesssim\mathcal{O}(1)$ when marginalizing over self-interaction templates, which indicates that (a) our analysis is probing qualitatively new regimes, and (b) there is much to be learned from future higher-resolution datasets. 

Finally, the technologies developed in this work can be straightforwardly applied to other primordial models. These could include additional strongly-coupled scenarios, such as the gapped unparticle model of \cite{Jiang:2025mlm}, the five-dimensional model of \citep{Kumar:2025anx}, the holographic model of \cite{Mishra:2025ofh}, or the more standard weakly-coupled collider \citep[e.g.,][]{Chen:2009zp,Arkani-Hamed:2015bza,Lee:2016vti}. While we have focused on de Sitter invariant systems in this work, this is also not a limitation: all parts of the pipeline extend trivially to inflationary models with strong breaking of scale-invariance. As shown in \cite{Pimentel:2025rds,Jiang:2025mlm}, unparticle exchange also yields novel signatures in the inflationary trispectrum; whilst a full-shape analysis along the lines of this work is not yet feasible, a simplified analysis focusing on factorizable signatures may be possible following \citep{Philcox4pt1,Philcox4pt3,2015arXiv150200635S}. We leave these extensions to future work.  

\vskip 8pt
\acknowledgments
{\small
\begingroup
\hypersetup{hidelinks}
\noindent 
This project arose from discussions at the 41st Annual Colloquium of the Institut d'Astrophysique de Paris. We thank Salvatore Sirletti for comments on the draft.
The computations in this work were run at facilities supported by the Scientific Computing Core at the Flatiron Institute, a division of the Simons Foundation. OHEP was inspired by the IAP's \href{https://www.flickr.com/photos/198816819@N07/55144647937/}{nauticall division}. 
GLP and CY are supported by Scuola Normale and by INFN (IS GSS-Pi). 
The research of GLP and CY is moreover supported by the ERC (NOTIMEFORCOSMO, 101126304). 
Views and opinions expressed are, however, those of the author(s) only and do not necessarily reflect those of the European Union or the European Research Council Executive Agency. 
Neither the European Union nor the granting authority can be held responsible for them. 
GLP is further supported by the Italian Ministry of Universities and Research (MUR) under contract 20223ANFHR (PRIN2022). 
\endgroup
\vskip 4pt
}

\appendix
\section{Constraints on Unparticles}

\setlength{\aboverulesep}{0pt}
\setlength{\belowrulesep}{0pt}
\setlength{\tabcolsep}{3pt}

\begin{longtable}{c||cc|cc @{\hspace{1cm}} c||cc|cc}
\caption{Unparticle non-Gaussianity constraints}\\
\cmidrule(r{1.1cm}){1-5} \cmidrule(l{-0.1cm}){6-10}
$\Delta$ & $f_{\rm NL}^{\rm unmarg}$ & $\sigma^{\rm unmarg}$ & $f_{\rm NL}^{\rm marg}$ & $\sigma^{\rm marg}$ & $\Delta$ & $f_{\rm NL}^{\rm unmarg}$ & $\sigma^{\rm unmarg}$ & $f_{\rm NL}^{\rm marg}$ & $\sigma^{\rm marg}$ \\
\cmidrule(r{1.1cm}){1-5} \cmidrule(l{-0.1cm}){6-10}
\endfirsthead

\caption{Unparticle non-Gaussianity constraints (continued)}\\
\cmidrule(r{1.2cm}){1-5} \cmidrule(l{-0.1cm}){6-10}
$\Delta$ & $f_{\rm NL}^{\rm unmarg}$ & $\sigma^{\rm unmarg}$ & $f_{\rm NL}^{\rm marg}$ & $\sigma^{\rm marg}$ & $\Delta$ & $f_{\rm NL}^{\rm unmarg}$ & $\sigma^{\rm unmarg}$ & $f_{\rm NL}^{\rm marg}$ & $\sigma^{\rm marg}$ \\
\cmidrule(r{1.2cm}){1-5} \cmidrule(l{-0.1cm}){6-10}
\endhead

\small
0.990 & 11.6 & 30.3 & -48.2 & 173 & 5.05 & -15.1 & 50.6 & 291 & 388 \\
1.05 & 11.6 & 30.3 & -47.9 & 173 & 5.10 & -15.3 & 52.4 & 328 & 510 \\
1.10 & 11.7 & 30.3 & -46.8 & 174 & 5.15 & -15.5 & 54.6 & 240 & 760 \\
1.15 & 11.8 & 30.3 & -44.9 & 177 & 5.20 & -15.6 & 57.2 & -897 & 1.03$\times10^3$ \\
1.20 & 12.0 & 30.5 & -42.2 & 183 & 5.25 & -15.3 & 60.1 & -666 & 575 \\
1.25 & 12.1 & 30.8 & -39.0 & 194 & 5.30 & -13.7 & 62.4 & -313 & 286 \\
1.30 & 12.3 & 31.3 & -35.0 & 209 & 5.35 & -8.82 & 60.2 & -164 & 156 \\
1.35 & 12.5 & 31.9 & -30.1 & 230 & 5.40 & 1.11$\times10^{-1}$ & 45.5 & -86.2 & 84.1 \\
1.40 & 12.6 & 32.6 & -23.8 & 257 & 5.45 & 3.84 & 22.5 & -38.8 & 38.5 \\
1.45 & 12.6 & 33.4 & -15.4 & 293 & 5.50 & 9.25$\times10^{-1}$ & 3.61 & -6.53 & 6.57 \\
1.50 & 12.5 & 34.3 & -4.15 & 338 & 5.55 & -2.54 & 8.80 & 17.3 & 17.6 \\
1.55 & 12.4 & 35.3 & 11.4 & 396 & 5.60 & -5.07 & 16.8 & 36.1 & 37.1 \\
1.60 & 12.2 & 36.2 & 33.1 & 470 & 5.65 & -6.86 & 22.4 & 51.8 & 53.6 \\
1.65 & 11.9 & 37.2 & 64.2 & 566 & 5.70 & -8.18 & 26.4 & 65.6 & 68.3 \\
1.70 & 11.6 & 38.2 & 110 & 692 & 5.75 & -9.23 & 29.6 & 78.4 & 82.1 \\
1.75 & 11.2 & 39.1 & 179 & 862 & 5.80 & -10.1 & 32.3 & 90.8 & 95.8 \\
1.80 & 10.8 & 40.0 & 291 & 1.10$\times10^3$ & 5.85 & -10.9 & 34.7 & 104 & 110 \\
1.85 & 10.3 & 40.8 & 486 & 1.44$\times10^3$ & 5.90 & -11.6 & 37.0 & 118 & 126 \\
1.90 & 9.83 & 41.6 & 867 & 1.99$\times10^3$ & 5.95 & -12.4 & 39.2 & 134 & 145 \\
1.95 & 9.36 & 42.3 & 1.77$\times10^3$ & 2.96$\times10^3$ & 5.99 & -13.0 & 41.1 & 149 & 163 \\
1.99 & 9.08 & 42.8 & 5.01$\times10^3$ & 4.89$\times10^3$ & 6.05 & -14.1 & 44.1 & 179 & 202 \\
2.05 & 8.53 & 43.5 & 1.64$\times10^4$ & 9.70$\times10^3$ & 6.10 & -15.1 & 47.0 & 215 & 253 \\
2.10 & 8.19 & 43.9 & 8.94$\times10^3$ & 7.41$\times10^3$ & 6.15 & -16.2 & 50.5 & 257 & 342 \\
2.15 & 8.04 & 44.1 & 2.11$\times10^3$ & 4.29$\times10^3$ & 6.20 & -17.6 & 54.7 & 186 & 498 \\
2.20 & 8.25 & 44.1 & 674 & 2.70$\times10^3$ & 6.25 & -19.0 & 59.7 & -332 & 517 \\
2.25 & 9.23 & 43.3 & 284 & 1.68$\times10^3$ & 6.30 & -19.8 & 64.9 & -268 & 287 \\
2.30 & 1.25 & 40.1 & 141 & 854 & 6.35 & -17.0 & 66.1 & -153 & 157 \\
2.35 & 9.11 & 7.68 & 11.4 & 53.0 & 6.40 & -7.51 & 54.5 & -88.7 & 91.0 \\
2.40 & -16.3 & 55.0 & -253 & 809 & 6.45 & 7.77$\times10^{-1}$ & 33.5 & -50.0 & 51.4 \\
2.45 & -6.90 & 52.8 & -815 & 1.77$\times10^3$ & 6.50 & 2.06 & 15.4 & -24.0 & 24.7 \\
2.50 & -4.55 & 52.0 & -1.86$\times10^3$ & 2.84$\times10^3$ & 6.55 & 5.80$\times10^{-1}$ & 3.02 & -5.06 & 5.23 \\
2.55 & -3.70 & 51.7 & -3.50$\times10^3$ & 3.95$\times10^3$ & 6.60 & -1.22 & 5.41 & 9.76 & 10.1 \\
2.60 & -3.38 & 51.6 & -5.68$\times10^3$ & 5.04$\times10^3$ & 6.65 & -2.80 & 11.4 & 22.1 & 22.9 \\
2.65 & -3.29 & 51.6 & -8.11$\times10^3$ & 6.06$\times10^3$ & 6.70 & -4.14 & 16.0 & 32.8 & 34.1 \\
2.70 & -3.30 & 51.7 & -1.03$\times10^4$ & 6.95$\times10^3$ & 6.75 & -5.29 & 19.7 & 42.8 & 44.5 \\
2.75 & -3.35 & 51.7 & -1.18$\times10^4$ & 7.61$\times10^3$ & 6.80 & -6.31 & 22.8 & 52.4 & 54.6 \\
2.80 & -3.41 & 51.8 & -1.18$\times10^4$ & 7.81$\times10^3$ & 6.85 & -7.27 & 25.6 & 62.2 & 64.9 \\
2.85 & -3.43 & 51.8 & -1.03$\times10^3$ & 7.32$\times10^3$ & 6.90 & -8.20 & 28.3 & 72.9 & 76.0 \\
2.90 & -3.41 & 51.9 & -7.87$\times10^3$ & 6.20$\times10^3$ & 6.95 & -9.16 & 31.0 & 85.1 & 88.8 \\
2.95 & -3.31 & 51.9 & -5.45$\times10^3$ & 4.84$\times10^3$ & 6.99 & -9.96 & 33.2 & 96.7 & 101 \\
2.99 & -3.16 & 51.9 & -3.92$\times10^3$ & 3.83$\times10^3$ & 7.05 & -11.3 & 36.8 & 119 & 125 \\
3.05 & -2.77 & 51.8 & -2.37$\times10^3$ & 2.60$\times10^3$ & 7.10 & -12.6 & 40.2 & 147 & 155 \\
3.10 & -2.21 & 51.7 & -1.55$\times10^3$ & 1.84$\times10^3$ & 7.15 & -14.3 & 44.4 & 187 & 202 \\
3.15 & -1.32 & 51.5 & -1.02$\times10^3$ & 1.29$\times10^3$ & 7.20 & -16.2 & 49.6 & 228 & 281 \\
3.20 & 9.04$\times10^{-2}$ & 51.0 & -668 & 875 & 7.25 & -18.7 & 56.1 & 73.1 & 356 \\
3.25 & 2.45 & 49.9 & -421 & 568 & 7.30 & -20.9 & 63.4 & -145 & 263 \\
3.30 & 6.64 & 47.0 & -245 & 337 & 7.35 & -19.9 & 67.6 & -117 & 152 \\
3.35 & 13.5 & 37.7 & -115 & 160 & 7.40 & -10.9 & 59.0 & -77.1 & 91.6 \\
3.40 & 6.69 & 8.20 & -16.9 & 23.7 & 7.45 & -1.34 & 39.8 & -49.2 & 56.3 \\
3.45 & -19.0 & 27.9 & 59.5 & 84.0 & 7.50 & 1.74 & 22.7 & -29.6 & 33.3 \\
3.50 & -20.6 & 42.4 & 121 & 170 & 7.55 & 1.51 & 10.7 & -15.1 & 16.8 \\
3.60 & -17.0 & 49.2 & 213 & 302 & 7.60 & 4.28$\times10^{-1}$ & 2.39 & -3.65 & 4.01 \\
3.70 & -15.0 & 50.9 & 283 & 405 & 7.65 & -7.41$\times10^{-1}$ & 3.63 & 5.93 & 6.47 \\
3.80 & -13.8 & 51.8 & 347 & 508 & 7.70 & -1.83 & 8.26 & 14.4 & 15.6 \\
3.90 & -13.0 & 52.6 & 416 & 647 & 7.75 & -2.82 & 12.0 & 22.2 & 23.9 \\
3.99 & -12.3 & 53.4 & 486 & 890 & 7.80 & -3.75 & 15.3 & 29.7 & 31.8 \\
4.10 & -11.2 & 54.6 & -575 & 2.03$\times10^3$ & 7.85 & -4.63 & 18.2 & 37.5 & 39.8 \\
4.20 & -9.25 & 56.0 & -1.26$\times10^3$ & 1.03$\times10^3$ & 7.90 & -5.51 & 21.0 & 45.8 & 48.3 \\
4.30 & -3.74 & 55.8 & -289 & 283 & 7.95 & -6.41 & 23.8 & 55.3 & 57.8 \\
4.40 & 8.87 & 31.5 & -652 & 68.1 & 7.99 & -7.18 & 26.1 & 64.4 & 66.7 \\
4.50 & -6.49 & 15.0 & 31.5 & 34.0 & 8.05 & -8.48 & 29.8 & 82.0 & 83.5 \\
4.60 & -12.2 & 32.6 & 87.7 & 96.6 & 8.15 & -11.3 & 37.8 & 136 & 133 \\
4.70 & -13.6 & 39.3 & 128 & 143 & 8.25 & -15.8 & 50.4 & 182 & 239 \\
4.80 & -14.1 & 43.1 & 163 & 187 & 8.35 & -18.0 & 65.2 & -69.4 & 140 \\
4.90 & -14.5 & 46.0 & 202 & 237 & 8.45 & -1.27 & 42.5 & -41.5 & 56.5 \\
4.99 & -14.8 & 48.6 & 248 & 307 & 8.55 & 1.98 & 15.3 & -18.0 & 22.3 \\
\end{longtable}

\bibliographystyle{apsrev4-1}
\bibliography{refs}

\end{document}